\documentclass[pra,twocolumn,superscriptaddress,longbibliography]{revtex4}
\usepackage{graphicx}
\usepackage{dcolumn}
\usepackage{bm}
\usepackage{physics}
\usepackage{amsmath}
\usepackage{array}
\usepackage{color}
\usepackage{times}
\newcolumntype{P}[1]{>{\centering\arraybackslash}p{#1}}
\usepackage[colorlinks=true, breaklinks=true, linkcolor=red, citecolor=blue, urlcolor=blue]{hyperref}

\newcommand{\bonnpi}{Physikalisches Institut, University of Bonn, Nussallee 12, 53115 Bonn, Germany}
\newcommand{\kl}{Physics Department and Research Center OPTIMAS, Technische Universit\"at Kaiserslautern, 67663 Kaiserslautern, Germany.}

\newcommand{\upd}[1]{^\mathrm{#1}}
\newcommand{\ind}[1]{_\mathrm{#1}}
\newcommand{\floor}[1]{\lfloor #1 \rfloor}
\newcommand{\ceil}[1]{\lceil #1 \rceil}

\begin{document}
\title{Momentum resolved Floquet-engineered pair and single particle filter in the Fermi Hubbard model}
\date{\today}
\begin{abstract}
We investigate the transport properties of a Fermi-Hubbard chain with an impurity which is formed by a site with a periodically modulated chemical potential. We determine the momentum resolved transmission through this impurity in dependence of the modulation frequency and strength for a single particle and a pair of fermions. We find that the pair transmission has a very distinct behaviour from the single particle transmission. Different situations can occur, where only the single particle or the pair with a certain momentum are transmitted or filtered out. 
\end{abstract}
\author{Friedrich H\"ubner}
\affiliation{\bonnpi}
\author{Christoph Dauer}
\affiliation{\kl}
\author{Sebastian Eggert}
\affiliation{\kl}
\author{Corinna Kollath}
\affiliation{\bonnpi}
\author{Ameneh Sheikhan}
\affiliation{\bonnpi}

\maketitle
\section{Introduction}
In the last decade a lot of advances were made in driving quantum systems in engineered ways. A particularly active field is the driving of systems by periodic modulations. Many interesting phenomena can be controlled by a periodic driving~\cite{EckardtAnisimovas2015}. One of the first phenomena demonstrated was the dynamical localization~\cite{LignierArimondo2007}. By the application of a linear potential which is periodically varied in time the particles in a lattice could be localized in an otherwise delocalized state. Further, artificial magnetic fields~\cite{AidelsburgerBloch2011,KollathBrennecke2016, SheikhanKollath2016, SheikhanKollath2016R, SheikhanKollath2019} could be engineered and  phase transitions could be driven~\cite{ZenesiniArimondo2009,PolettiKollath2011,KitamuraAoki2016,FazziniEggert2021,SheikhanKollath2020} or the transport of bound pairs could be controlled \cite{KudoMonteiro2009}.

A particularity of time-periodically modulated systems is that the so-called Floquet theory~(see review \cite{EckardtAnisimovas2015} and references therein), can be devised which uses a similar idea as the Bloch's theorem which describes systems which are periodic in space. The time-dependent Schr\"odinger equation is reformulated in terms of an eigenvalue problem with so-called quasi-energies which correspond to the quasi-momenta in the spatial periodic structure.  

Here we consider a periodically driven impurity and its interplay with the interactions in a quantum system. The system under consideration is a Fermi-Hubbard chain in which a single site has a time-periodically modulated chemical potential. This is the impurity site. Systems with a local periodic driving have been studied previously and complex behaviour has been found. In the continuum, the scattering properties of a single particle at a periodically driven $\delta$ potential have been investigated \cite{MartinezReichl2001}.  A lot of interesting predictions have been made for the transmission properties of a single particle through this 'quantum dot'  \cite{ReyesEggert2017,ThubergEggert2016} or through the quantum point contact \cite{GamayunLychkovskiy2021}.  Further, a periodically modulated quantum dot has been proposed to induce bound states \cite{AgarwalaSen2017} or as a spin filter~\cite{ThubergReyes2017} for which one spin species is fully blocked while the other is fully transmitted by the quantum dot. 

In many of these studies an important effect, the interaction between the fermions, has been neglected. Here we investigate specifically the interplay between the interaction and the driving. We compare the transport behaviour of a single particle (without interactions) and a pair of particles which is strongly bound by the attractive interaction in the system. By the driving a very different behaviour between the two transmission can be found. This work extends our previous work \cite{HuebnerSheikhan2022}, where in the same setup the momentum integrated transmission was discussed. In this work we analyze in detail the dependence on the momentum resolved transmission behaviour and its special features.

In section \ref{sec:model} we describe the setup. In section \ref{sec:single} we summarize for completeness the results on the previously studied single particle transport. In section \ref{sec:pair} we discuss the momentum dependent transmission of pairs through the driven impurity. In the limit of low driving frequency (subsection \ref{sec:pair_low}), a Schrieffer-Wolff transformation leads to a Hamiltonian which is similar to the single particle case with effective parameters. Additionally, in the limit where the driving and the interaction compete on similar energy scales (subsection \ref{sec:pair_high} and following) we use the Floquet-Schrieffer Wolff transformation in order to obtain the momentum resolved transmission and analyze its behaviour. In section \ref{sec:pair_discussion} we discuss the results of the pair transmission. In section \ref{sec:filter}, different ways how to use the driven impurity as filters are outlined. 

\section{Model}
\label{sec:model}
\begin{figure}[hbtp]
\centering
\includegraphics[width=.4\textwidth]{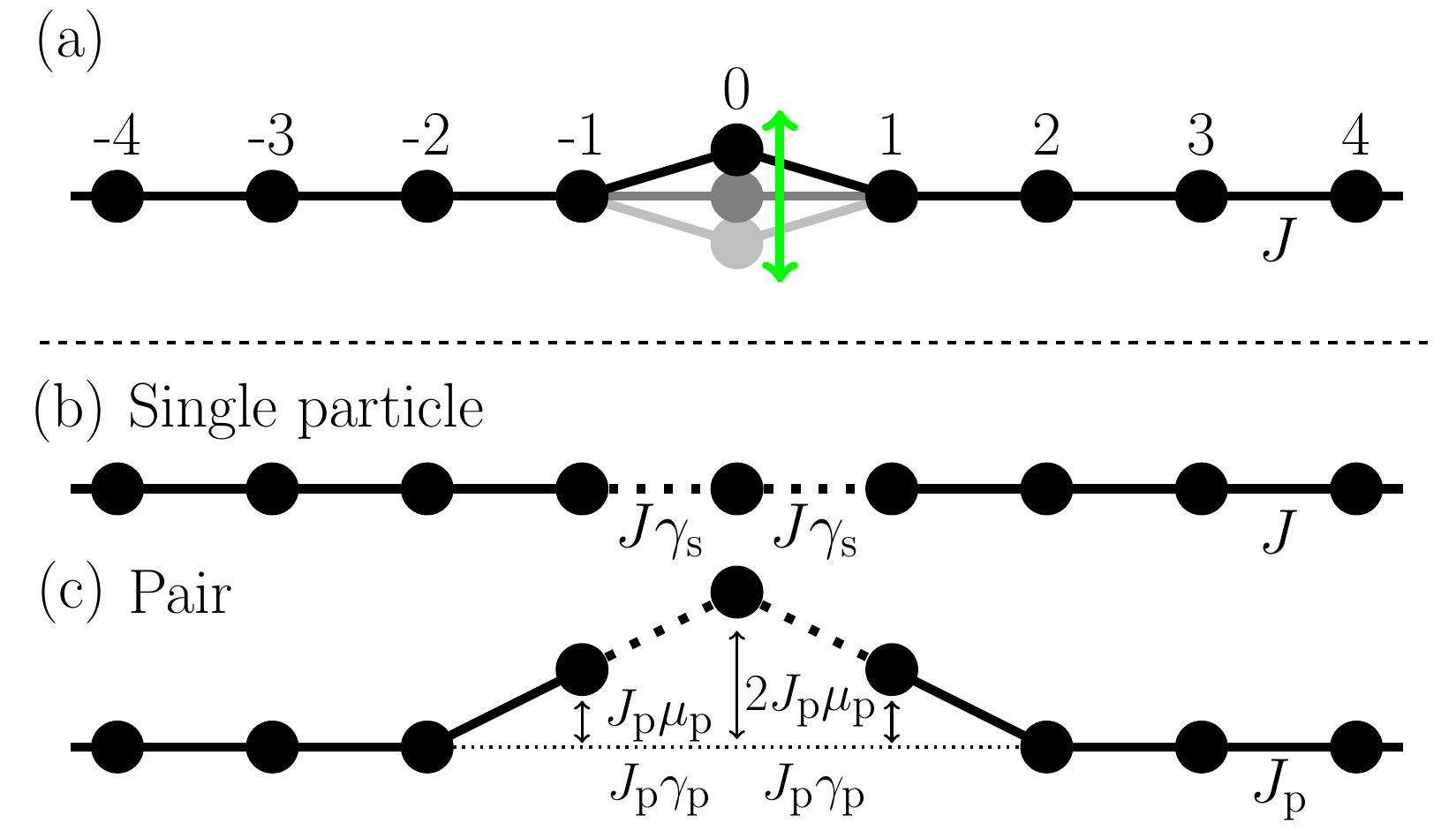}
\caption{(a) Sketch of the Fermi-Hubbard chain with a periodically driven site in the middle. $J$ denotes the hopping amplitude on the bonds connected with solid lines.  (b) Sketch of the effective model in the high frequency limit for a single particle.  The amplitude of the hopping $J$ is represented by the solid lines and the effective hopping amplitudes (dotted lines) to and from the central site are modified to $\gamma_sJ$. (c) Sketch of the effective model for a particle pair. The effective hopping amplitude (solid lines) away of the center is given by $J_p$. The effective hopping amplitudes (dotted lines) around the central site are modified to $\gamma_p J_p$. In contrast to the single particle case, an additional chemical potential shift exists at the three central sites. }
\label{fig:setup}
\end{figure}
We consider a Fermi-Hubbard chain with an impurity given by an oscillating chemical potential at the impurity site  (see Fig. \ref{fig:setup} a). The corresponding Hamiltonian is, 

\begin{align}
\vb{H}(t) = -J&\sum_{n,\sigma} \qty(\vb{c}_{n\sigma}^\dagger\vb{c}_{n+1\sigma} + \mathrm{h.c.}) + U\sum_n \vb{n}_{n\uparrow}\vb{n}_{n\downarrow}\nonumber\\
 &+ \hbar\lambda\omega\cos(\omega t) \qty(\vb{n}_{0\uparrow} + \vb{n}_{0\downarrow}).
\label{equ:hamiltonian}
\end{align}

Here $\vb{c}_{n\sigma}$ is the annihilation operator for a fermion of spin $\sigma$ at site $n$, $\vb{n}_{n\sigma} = \vb{c}_{n\sigma}^\dagger\vb{c}_{n\sigma}$, $J$ is the hopping parameter, $U < 0$ the attractive onsite Hubbard interaction. The impurity is located at site $n=0$. The driving frequency is denoted by $\omega$. The amplitude of the external perturbation is the driving strength $\lambda$ in units of $\omega$. Note, we fixed the lattice spacing to $a = 1$ and thus all appearing momenta will be dimensionless, i.e.~in units of $1/a$. Further we use $\hbar=1$. By performing a gauge transformation with the unitary operator $\vb{U}(t) = e^{-i\lambda(\vb{n}_{0\uparrow} + \vb{n}_{0\downarrow})\sin{\omega t}}$ one gets a new time-dependent Hamiltonian, 
\begin{align}
\vb{H}\upd{g}(t) = -J&\sum_{n\sigma} \qty(g_n(\omega t)\vb{c}_{n\sigma}^\dagger\vb{c}_{n+1\sigma} + \mathrm{h.c.}) + U\sum_n \vb{n}_{n\uparrow}\vb{n}_{n\downarrow}, \label{equ:hamiltonian_gauge}
\end{align}
where $g_n(\phi) = 1$ unless $n = -1,0$, where $g_{-1}(\phi) = e^{-i\lambda \sin(\phi)}$ and $g_{0}(\phi) = e^{i\lambda \sin(\phi)}$. This gauge transformation introduces a time-dependent hopping at bonds connected to the impurity site  instead of the driving chemical potential at the impurity.  Note that after the gauge transformation the impurity does not only act on site $0$, but rather on the bonds connecting the three sites $n= -1 ,0$ and $1$. 

\section{Single particle transport}
\label{sec:single}
The single particle transport in this situation has been studied in detail in Ref.~\cite{ThubergEggert2016}. For completeness we summarize some of the results here in order to contrast them to the two-particle behaviour. In the high frequency limit $\omega \gg J$, one can use the usual high-frequency expansion~\cite{TakegoshiMadhu2015,EckardtAnisimovas2015} to derive an effective static Hamiltonian by averaging over a full period $T = 2\pi/\omega$:

\begin{align}
  \vb{H}\upd{s} =& \int_0^{T} \vb{H}\upd{g}(t) \frac{\dd{t}}{T}\\ 
 = &-J\sum_{n\neq -1,0} (\vb{c}_{n\uparrow}^\dagger\vb{c}_{n+1\uparrow} + \mathrm{h.c.})\nonumber \\
&- J\gamma_s(\lambda)\sum_{n = -1,0} (\vb{c}_{n\uparrow}^\dagger\vb{c}_{n+1\uparrow} + \mathrm{h.c.}).
\label{equ:hamiltonian_single}
\end{align}

Here $\gamma\ind{s} = J_0(\lambda)$ is the zero'th Bessel function. A sketch of this effective model is drawn in Fig.~\ref{fig:setup} b). It is simply a chain with a static impurity, given by a rescaled hopping amplitude from and to the impurity at site $n=0$. In the effective static model the transmission of a single particle can be calculated analytically by matching the wave functions in the different (constant) regions. The single particle transmission in the high frequency limit is given by
\begin{align}
	T\upd{s}_k = \frac{1}{1 + \qty(\frac{1}{\gamma\ind{s}^2}-1)^2\cot^2 k}.
	\label{equ:transmission_single}
\end{align}

\begin{figure}[hbtp]
\centering
\includegraphics[width=.48\textwidth]{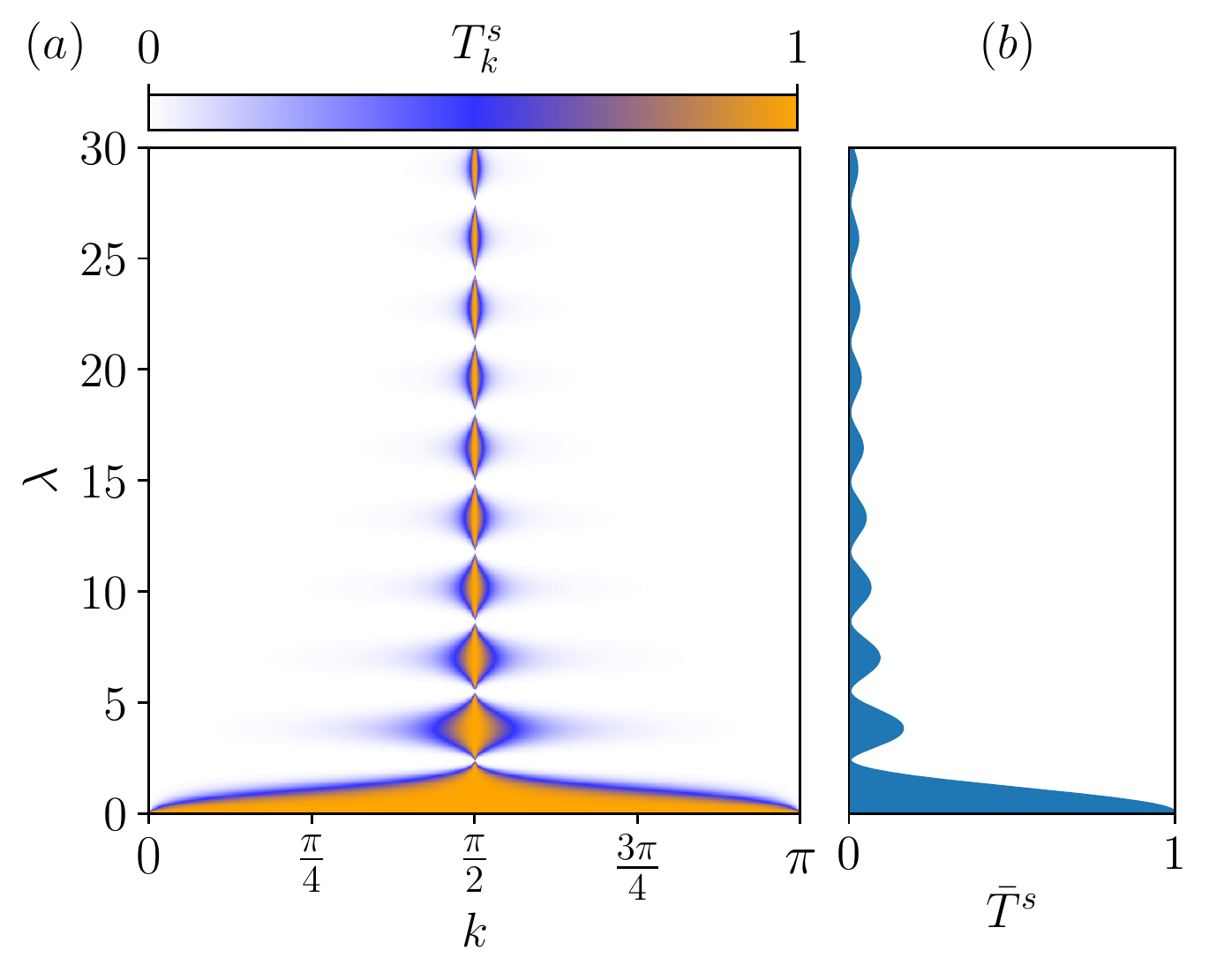}
\caption{Single particle transmission $T^s_k$ as a function of the momentum $k$ and driving strength $\lambda$ in the high frequency limit ($\omega \gg J$). The plot on the right side shows the momentum averaged transmission $\bar{T}^s$ versus $\lambda$. }
\label{fig:single_transmission}
\end{figure}

In Fig.~\ref{fig:single_transmission} the momentum resolved single particle transmission is plotted. For a fixed driving strength $\lambda$ and large driving frequencies, the transmission can be described as follows. The transmission shows a symmetric peak with a maximum value at momentum $k = \pi/2$. At this special value of the momentum $k=\pi/2$ which corresponds to zero energy in the single particle dispersion, the particle is always fully transmitted with $T\upd{s}_{k} = 1$ for $k = \pi/2$. For $k = 0$ or $k = \pi$, i.e.~ at the center or the boundaries of the Brioullin zone, the transmission vanishes for finite driving. The width of the peak around the momentum $k=\pi/2$ is determined by the Besselfunction $J_0(\lambda)$. Thus, there is full transmission for all momenta when $J_0(\lambda) = 1$, which only happens for $\lambda = 0$. The transmission is completely suppressed at a driving which corresponds to the zeros of the Bessel function $J_0(\lambda) = 0$. The first of these zeros occur at $\lambda = 2.4, 5.52, 8.65, \ldots$. For completeness, the momentum integrated transmission is also represented in Fig.~\ref{fig:single_transmission} and shows the oscillating behaviour. Thus, one can always easily suppress any transmission of single particles by choosing $\lambda$ as one of these values. 

\begin{figure}[hbtp]
\centering
\includegraphics[width=.4\textwidth]{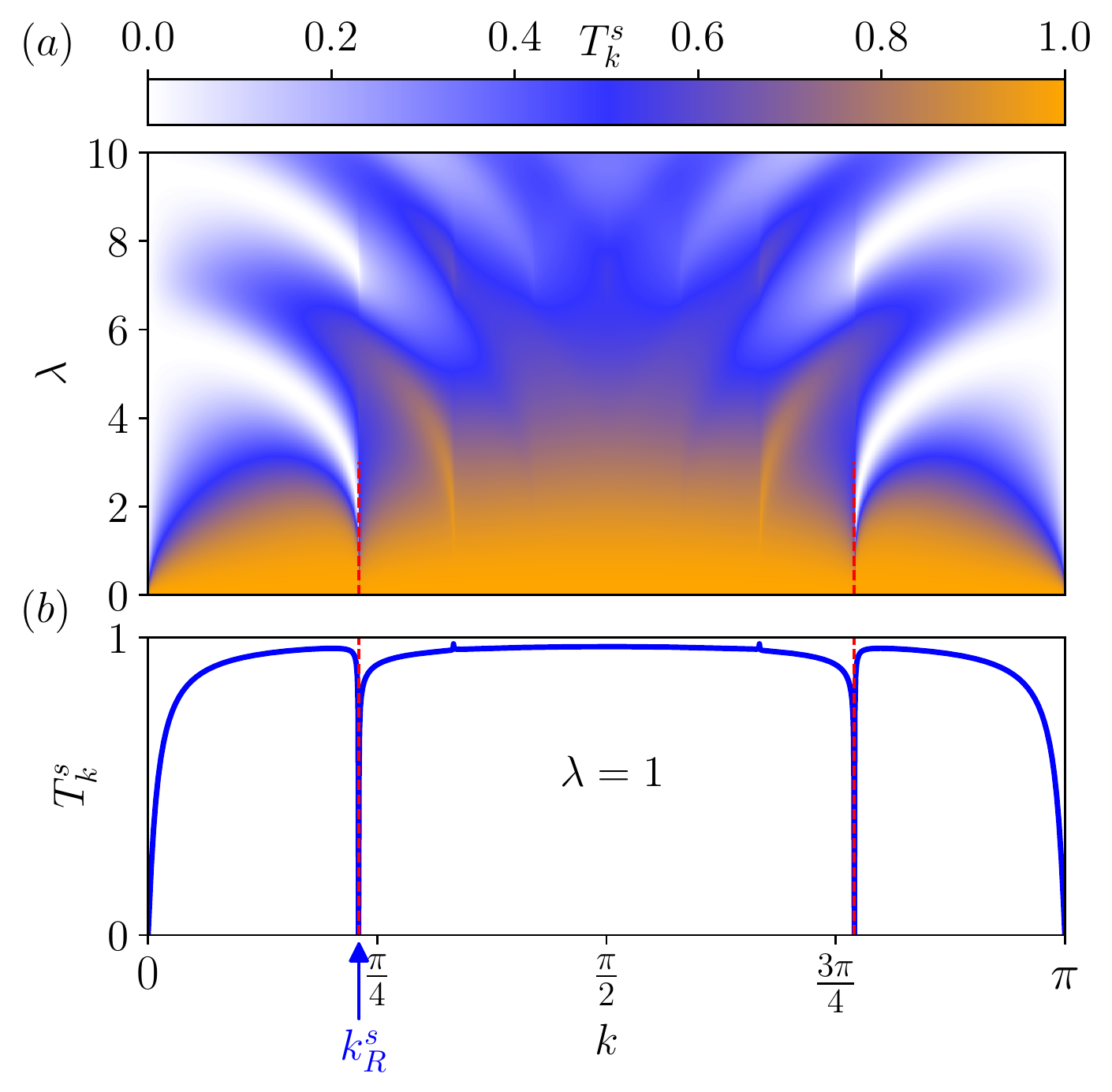}
\caption{a) Single particle transmission $T^s_k$ as a function of momentum $k$ and driving strength $\lambda$ for a low frequency $\omega = 0.5J$. The clear Fano-resonances at which the transmission vanishes are seen at $k_r$ which are marked by a red dashed line.  b) A cut of the transmission is shown for $\lambda= 1$.}
\label{fig:single_transmission_fano}
\end{figure}

A more complex phenomena can be found for the single particle transmission at low frequency $\omega \approx J$ as shown in Fig.\ref{fig:single_transmission_fano}.  In this case pronounced resonances occur for certain momenta at which the transmission vanishes \cite{ThubergEggert2016}. These resonances are associated with so-called Fano resonances. They occur due to the scattering at bound states which occur in different scattering channels, here the different frequency side bands. In the vicinity of the transmission zero the transmission coefficient can be approximately described by the formula $T_s=\alpha_k^2/(\alpha_k^2+1)$, with $\alpha_k=2 J \sin(k) [\epsilon_{\rm R}+2 J \cos (k)]/\Delta_{\rm R}$. The resonance energy $\epsilon_{\rm R}^s$ for a single particle  behaves as
\begin{equation}
\label{eq:fano_single}
\epsilon_{\rm R}^s=\pm[2 J-\omega+\lambda^4\omega^3/(64 J(4J+\omega))]
\end{equation}
for small $\lambda \omega$\cite{ReyesEggert2017}, while the width $\Delta_{\rm R}$ of the feature scales as $\Delta_{\rm R} \propto \lambda^4$ for small $\lambda \omega$. The momenta at which the resonances occur are then given by the relation $\epsilon_{\rm R}^s=2J_p\cos(k_{\rm R}^s) $.

These resonances can be employed to design a momentum sensitive filter suppressing the transmission at certain momenta. For example, by changing the driving frequency $\omega$ the position of the Fano resonance, i.e.~the momentum $k_R^s$, at which the Transmission zero occurs, can be chosen. The range of suppressed momenta, which corresponds to the width $\Delta_{\rm R}$, increases with $\lambda \omega$.

\section{Pair Transport}
\label{sec:pair}
For large attractive interaction in the static Hubbard chain one can use the Schrieffer-Wolff transformation to derive an effective Hamiltonian for pairs of fermions which includes one fermion with spin up and one with spin down.

However, in the presence of the driven impurity, another energy scale enters into the play which is the driving frequency. Thus, one has to distinguish two different cases, the low frequency case, where the driving $\omega \ll |U|$ is much lower than the strong interaction strength and the limit where both energy scales become comparable. Whereas in the first case, the well known Schrieffer-Wolff transformation can be applied, for the second case more care has to be taken and  we need to apply an extension of the Schrieffer-Wolff transformation for periodically-driven systems \cite{BukovPolkovnikov2016}.

\subsection{Schrieffer-Wolff transformation in the limit of low driving frequency}
\label{sec:pair_low}
In this section we consider the case of strong attractive interaction, but low driving frequency, i.e.~$J, \omega \ll |U|$. At such low frequencies, the pairs are not broken by the driving and we can employ a Schrieffer-Wolff transformation which maps the system onto a model of stable onsite pairs. The effective pair Hamiltonian in this case is given by the time-dependent Hamiltonian
\begin{align}
\vb{H}\upd{eff}(t) = -J\ind{p}\sum_{n} \qty(\bm{\eta}_n^+ \bm{\eta}_{n+1}^- + \mathrm{h.c.}) + 2\lambda \omega\cos(\omega t) \bm{\eta}_0^+\bm{\eta}_0^-.
\label{eq:effpairs}
\end{align}
The newly introduced operators $\bm{\eta}_n^+ = \vb{c}_{n\uparrow}^\dagger\vb{c}_{n\downarrow}^\dagger$ and $\bm{\eta}_{n+1}^- = \vb{c}_{n\downarrow}\vb{c}_{n\uparrow}$ are pair `creation' and `annihilation' operators.

This effective time-dependent pair Hamiltonian $\vb{H}\upd{eff}(t)$ has the same form as the initial Hamiltonian Eq.~\ref{equ:hamiltonian}, however, with effective parameters. The effective tunneling amplitude is given by $J_p=\frac{2J^2}{|U|}$ and the driving amplitude of the impurity is $2\lambda$. The enhanced driving amplitude can be explained by the presence of two particles within each pair which feel twice as strongly the impurity potential. 
Due to the form of the Hamiltonian,  the scattering properties of the effective model $\vb{H}\upd{eff}(t)$ resemble the known behaviour for the single particles exchanging the parameters by the effective parameters.

In particular, for large frequencies $J \ll \omega \ll |U|$ the pair transmission is given by the expression in (\ref{equ:transmission_single}), only with the replacement $\gamma\ind{s} \to \gamma\ind{p} = J_0(2\lambda)$, i.e.~
\begin{align}
	T\upd{p}_k = \frac{1}{1 + \qty(\frac{1}{\gamma\ind{p}^2}-1)^2\cot^2 k}.
	\label{equ:integrated_pair_highfrequency}
\end{align}

Furthermore, at low frequencies which are comparable to $\omega \sim J_{\rm P} \ll J$, Fano-resonances occur in the transmission of the pair at momenta $k_{\rm R}$ with energy $\epsilon_{\rm R}^p$  given by the relation
\begin{equation}
\label{eq:fano}
  \epsilon_{\rm R}^p=2J_p\cos(k_{\rm R}^p) =\pm[2 J_{\rm P}-\omega+\lambda^4\omega^3/(4 J_{\rm P}(4J_{\rm P}+\omega))].
  \end{equation}

\begin{figure}[hbtp]
\centering
\includegraphics[width=.4\textwidth]{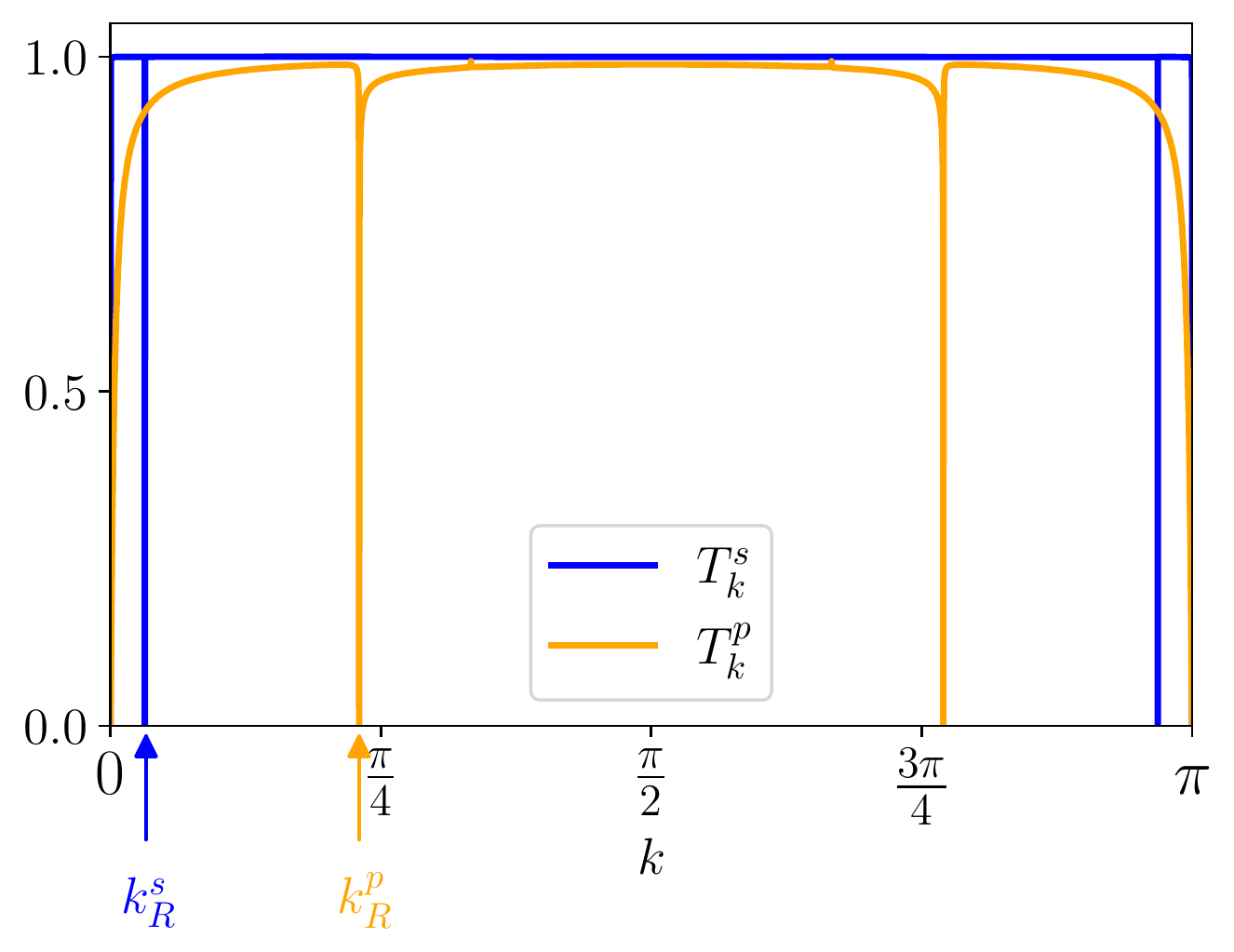}
\caption{Pair and single particle transmission as function of momentum $k$ for the driving strength $\lambda=0.3$, interaction strength $U=-100J$ and for a low frequency $\omega =0.01 J= 0.5J_p$. Clear resonances are seen in the pair transmission at approximately $-2J_p\cos(k_{\rm R}^p) =\pm (2 J_{\rm P}-\omega)$ (lowest order of Eq.\ref{eq:fano}) and for single particle transmission at $-2J\cos(k_{\rm R}^s) =\pm (2 J-\omega)$ where one of the resonances is marked by an arrow. At the resonant momentum of the pair where $T^p_{k_{\rm R}^p}\approx 0$ the impurity is almost transparent for the single particles, i.e.~$T^s_{k_{\rm R}^p}\approx 1$.}
\label{fig:singlepair_transmission_fano}
\end{figure}

This makes it possible to create momentum dependent filters for the pairs. Such a situation in which the impurity filters out certain momenta of the pair is a shown in Fig.~ \ref{fig:singlepair_transmission_fano}. The Fano resonances are seen as clear dips in the pair transmission. At the same momenta, the impurity is almost transparent for the single particles as $\omega\ll J$. 

\subsection{Floquet-Schrieffer-Wolff transformation}
\label{sec:pair_high}
For larger frequencies we have to consider the complex interplay with the interaction.  To describe this situation, we consider the limit of both large frequencies and large interactions $|U|,\omega \gg J$.  In this case it is possible to use the so-called Floquet-Schrieffer-Wolff transformation, which  performs the Schrieffer-Wolff transformation in the extended Hilbert space as described in the Appendix \ref{app:fswt}. Applying this Floquet-Schrieffer-Wolff transformation to the considered setup, the resulting effective pair Hamiltonian looks as follows:
\begin{align}
\vb{H}\upd{p} = -J\ind{p}\Bigl[&\sum_{n\neq -1,0} \qty(\bm{\eta}_n^+ \bm{\eta}_{n+1}^- + \mathrm{h.c.})\nonumber\\ 
&+ \gamma\ind{p} \sum_{n = -1,0} \qty(\bm{\eta}_n^+ \bm{\eta}_{n+1}^- + \mathrm{h.c.})\nonumber\\ 
&+ \mu\ind{p} (\vb{n}_{-1\uparrow} + 2\vb{n}_{0\uparrow} + \vb{n}_{1\uparrow})\Bigr].
\label{equ:hamiltonian_pair}
\end{align}

Thus, as for the single particle an effective static Hamiltonian is recovered. The effective model for a single pair is a chain with a hopping amplitude $J_p=2\frac{J^2}{|U|}$ on most of the bonds. Only around the bonds connecting to the impurity site $n=0$ are scaled by $\gamma\ind{p}$. In contrast to the single particle case,  the effective pair model contains also an extra triangular potential at sites $n=-1, 0$ and $1$. At site $n=\pm1$ the additional potential is given  by $\mu\ind{p}$. The value of the chemical potential at impurity is twice the neighbouring sites, i.e.~$2\mu_p$. The scaling of the reduced hopping and the Floquet induced chemical potential are given by

\begin{align}
\gamma\ind{p} &= \frac{|U|}{\omega}\sum_l \frac{ (-1)^l J_l(\lambda)^2}{\frac{|U|}{\omega}-l}\label{equ:gamma}\\
\mu\ind{p} &= \frac{|U|}{\omega}\sum_l \frac{ J_l(\lambda)^2}{\frac{|U|}{\omega}-l} - 1\label{equ:mu}.
\end{align}

Here $J_l(\lambda)$ is the $l$'th Bessel function. Note that the effective model is only valid if $|U|$ is not an integer multiple of $\omega$. In that case there would be the possibility of a pair breaking into two single particles at the impurity which cannot be captured by the effective pair Hamiltonian \footnote{When going away from the limit $|U|,\omega \gg J$ one finds that $4 J/\omega$ should be below the distance of $|U|/\omega$ to the nearest integer in order to avoid pair breaking. For example for a half-integer $|U|/\omega$ the maximum possible value of $J/\omega$ is $J/\omega = 0.125$. If $|U|/\omega$ is even closer to an integer, the maximum value of $J/\omega$ will be even smaller.}. However, using the Lippmann-Schwinger formalism it is still possible to treat the problem analytically in the regime $J \ll \omega,|U|$~\cite{Huebner2022}.

\subsection{Dependence of the effective parameters on the driving parameters}
We will discuss in this subsection, the dependence of the effective parameters $\gamma\ind{p}$  and $\mu\ind{p}$ on $\lambda$ and $|U|/\omega$. The expression for the effective parameters (Eq.~\ref{equ:gamma} and Eq.~\ref{equ:mu}) are given as infinite sums over Bessel functions and a denominator, which changes considerably the weight of the different contributions and can lead to pronounced maxima.
Whereas both expressions look very similar, the difference is that the extra $(-1)^l$ factor in $\gamma\ind{p}$ and the constant shift by $-1$ of $\mu_p$.

\begin{figure}[!h]
	\centering
	\includegraphics[width=.48\textwidth]{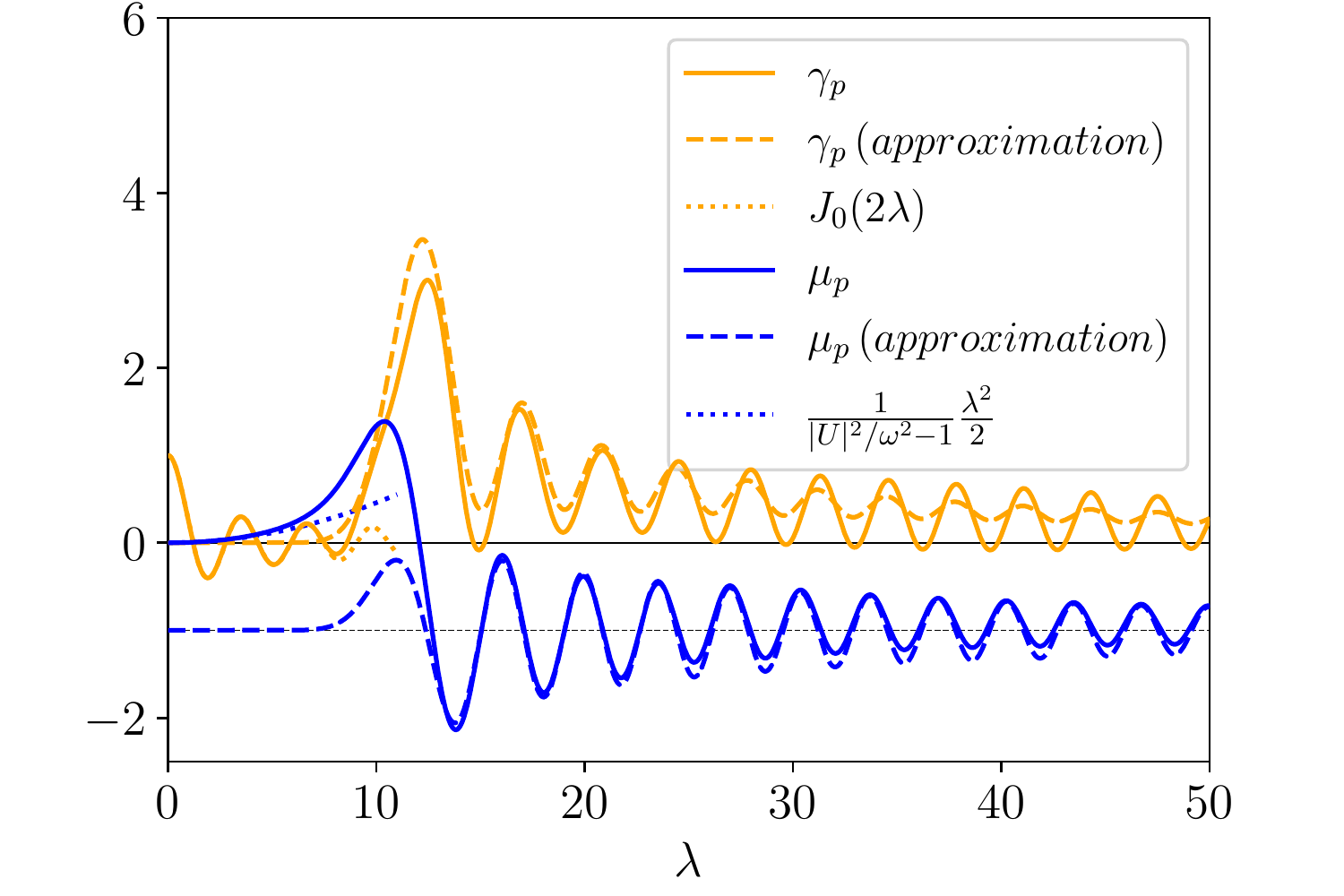}
	\caption{The parameters $\gamma\ind{p}$ and $\mu\ind{p}$ as function of driving strength $\lambda$ for $|U|/\omega = 10.5$. The dashed lines show the approximations (\ref{equ:gamma_approx}) and (\ref{equ:mu_approx}) where the infinite sums are approximated by two terms. The dotted lines give approximations (\ref{equ:gamma_taylor}) and (\ref{equ:mu_taylor}) which are valid for small $\lambda$.}
	\label{fig:params_gammamu_U105}
\end{figure}

We show a typical behaviour of the effective parameters in Fig.~\ref{fig:params_gammamu_U105}. The function $\gamma_p$ starts at the value $1$ for $\lambda=0$ and then oscillates around zero until it reaches a maximum close to the value $\lambda\approx |U|/\omega$. The maximal value of $\gamma_p$ can be large. For larger values of $\lambda$ the function $\gamma_p$ decays in an oscillatory manner towards zero. In contrast, the effective chemical potential $\mu_p$ behaves differently. At $\lambda=0$ its starts from zero and then rises to a maximum which occurs slightly below $\lambda\approx |U|/\omega$. For larger values of $\lambda$, the function $\mu_p$ rapidly decreases to a negative value around which it oscillates for larger values of $\lambda$.

Let us analyse this behaviour in the following in detail. The Bessel functions $J_l(\lambda)$ are oscillating functions which decay as $1/\sqrt{\lambda}$ and are thus all of comparable size. Therefore, the size of the terms is mainly determined by the prefactor $\frac{1}{\frac{|U|}{\omega} - l}$. The dominant contributions stem from the terms where the prefactor is maximal, i.e. for the two integers $l_- = \floor{|U|/\omega}, l_+ = \ceil{|U|/\omega}$ closest to $|U|/\omega$. Thus, we approximate $\gamma\ind{p}$ and $\mu\ind{p}$ only considering these two dominating terms,

\begin{align}
\gamma\ind{p} &\approx\frac{|U|}{\omega}\left( \frac{ (-1)^{l_-} J_{l_-}(\lambda)^2}{\frac{|U|}{\omega} - l_-} + \frac{(-1)^{l_+} J_{l_+}(\lambda)^2}{\frac{|U|}{\omega} - l_+}\right)\label{equ:gamma_approx}\\
\mu\ind{p} &\approx \frac{|U|}{\omega}  \left(\frac{J_{l_-}(\lambda)^2}{\frac{|U|}{\omega} - l_-} + \frac{\frac{|U|}{\omega} J_{l_+}(\lambda)^2}{\frac{|U|}{\omega} - l_+} - 1 \right). \label{equ:mu_approx}
\end{align}

To access the quality of this approximation, in Fig. \ref{fig:params_gammamu_U105} the exact expression of $\gamma\ind{p}$ and $\mu\ind{p}$ together with the above approximations is plotted as a function of the driving strength $\lambda$ for a fixed value of interaction. For the function $\gamma\ind{p}$ one can observe that for $\lambda > |U|/\omega$ the approximation is able to describe some of the important features, like the approximate position of the maxima and the frequency of oscillations. The approximation, however, fails to obtain the amplitude of oscillations which differ considerably from the exact value and the behaviour at low $\lambda$, where the approximate function just vanishes. This is due to the behaviour of the Bessel functions $J_{l_\pm}(\lambda)$ which basically vanish at low $\lambda$ and the first maxima appears around $\lambda \approx l_\pm$. Thus the sums in Eq.~\ref{equ:gamma} are dominated by other terms with indices $l<l_{-}$ which are not considered in the approximation in Eq. \ref{equ:gamma_approx}.
 
A similar behaviour is seen for the effective chemical potential $\mu\ind{p}$. At large values of $\lambda > |U|/\omega$ the approximation is able to describe the function relatively well. However, the behaviour at small $\lambda$ up to the first rise of the function is not at all covered by the approximation, since also here the contributions of the Bessel functions $J_{l_\pm}(\lambda)$ become subdominant. 

For low values of $\lambda$, the sum over the Bessel function is dominated by the lowest values of $l$, since the first maximum of the Bessel function is moving to larger values with larger $l$. Thus, in the limit $\lambda \ll |U|/\omega$, we can approximate the denominator $\frac{|U|}{\omega}-l$ neglecting the contribution of $l$, such that the dependence on $|U|/\omega$ drops out and we obtain the analytic expressions:
\begin{align}
\gamma\ind{p} &\to \sum_l (-1)^l J_l(\lambda)^2 = J_0(2\lambda)\label{equ:gamma_taylor}\\
\mu\ind{p} &\to \sum_l J_l(\lambda)^2 - 1 = 0.
\end{align}    
Here we used Neumann's addition theorem (see~\cite{NIST:DLMF}~\S 10.23). For small driving strength $\lambda$ we can also compute the leading corrections to $\mu_p=0$:
\begin{align}
\mu\ind{p} = \frac{1}{\frac{|U|^2}{\omega^2} - 1} \frac{\lambda^2}{2} + \order{\lambda^4},
\label{equ:mu_taylor}
\end{align}
which shows that for $|U|/\omega > 1$ and small $\lambda \ll |U|/\omega$, $\mu\ind{p}$ increases with increasing $\lambda$. The approximations are shown in Fig.~\ref{fig:params_gammamu_U105} and we see that these cover much better the values at low $\lambda$.

The low frequency expressions $\gamma\ind{p} = J_0(2\lambda)$ and $\mu\ind{p} = 0$ in the regime $J \ll \omega, |U|$ match to the high frequency results in the regime $J, \omega \ll |U|$ for which we derived the effective Hamiltonian (\ref{eq:effpairs}). In particular, since a pair consists of two particles, the argument of Bessel function is $2\lambda$ instead of $\lambda$.

\subsection{Determining the transmission for the effective pair Hamiltonian}
\label{sec:transmission_effective_pair_hamiltonian}
Using the effective parameters, one can compute the pair transmission at the impurity from the effective time-independent pair Hamiltonian (Eq.~\ref{equ:hamiltonian_pair}) \cite{HuebnerSheikhan2022}. As shown in Fig.\ref{fig:setup} (c) in case of pair transport the impurity in the effective model includes three sites in the center. One makes an ansatz for the wavefunction with incoming, transmitted, and reflected waves with amplitudes 1,$r_k$ and $t_r$, respectively considering that the three sites in the center of the chain are different. This ansatz reduces to an ansatz which contains only three different regions described by:
\begin{align}
\psi_n = \begin{cases} 
      e^{ikn} + r_k e^{-ikn} & n<0\\
      \psi_0 & n=0 \\
      t_k e^{ikn} & n>0
   \end{cases}.
   \label{equ:app_trans_wavefunction}
\end{align}

Inserting this ansatz into the Schr\"odinger equation with the Hamiltonian given in Eq.~\ref{equ:hamiltonian_pair}), and solving for $t_k$ yields
\begin{align}
t_k &= \frac{\gamma\ind{p}^2}{1-\mu\ind{p}e^{ik}}\frac{-2i\sin k}{\qty(2\cos k-2\mu\ind{p})\qty(1-\mu\ind{p} e^{ik})-2\gamma\ind{p}^2e^{ik}}.
\end{align}

From this expression one can compute the momentum dependent pair transmission $T_k = |t_k|^2$

\begin{align}
	T\upd{p}_k =& \frac{\gamma_p^4}{1 + \qty(\frac{\cos{k}-\mu_p}{\sin{k}})^2}\times\nonumber\\
	&\frac{1}{\qty[\qty(\cos{k}-\mu_p)^2-\gamma_p^2]^2 +\qty(\cos{k}-\mu_p)^2\sin^2{k}}.
	\label{equ:transmission_pair}
\end{align}

\begin{figure*}
\setlength{\unitlength}{0.1\textwidth}
\includegraphics[width=.99\textwidth]{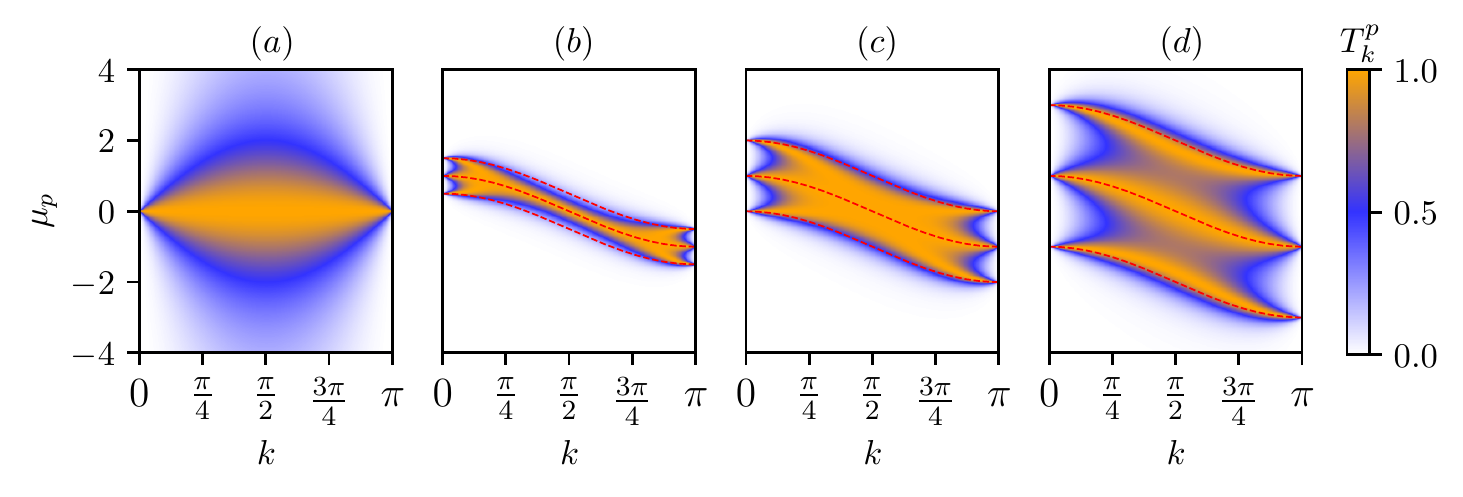}
\caption{ Discussion of the pair transmission versus the momentum and its dependence on the effective parameter $\mu_p$ and $\gamma_p$. (a) The case of a single effective potential $\mu_p$  on the middle site. The pair transmission for the triangular shaped effective potential of the pair model with the effective hopping (b) $\gamma_p=0.5$, (c) $\gamma_p=1$ and (d) $\gamma_p=2$. The red dashed lines in (b)-(d) are $\cos(k)-\mu_p\approx 0, \pm \gamma_p$. }
\label{fig:pairtransmission_mu_gamma}
\end{figure*}

The expression for the pair transmission is much more complicated than the single particle transmission Eq.\ref{equ:transmission_single}. We verified that the single particle expression can be recovered setting $\mu_p=0$ and $\gamma_p=\gamma_s$. In appendix \ref{app:pair} we show that this function can be written with just two parameters and analyze its dependence. In the following for a better understanding we point out several feature which arise due to the form of the effective potential. As expected, the presence of the potential not only on site 0 but also on site $\pm 1$ causes a more complex structure of the transmission versus momentum as is shown in Fig.~\ref{fig:pairtransmission_mu_gamma}. Whereas in the single impurity case the transmission has a symmetry around the momentum $\pi/2$ and under a sign change of the potential $\mu_p \to -\mu_p$, only a combined symmetry where $\cos(k)-\mu_p$ changes sign remains in the triangular shaped potential. The transmission is large for positive $\mu_p$ and momenta close to $k=0$ and for negative $\mu_p$ and momenta close to $k=\pi$. A clear substructure occurs due to the triangular shape of the effective pair potential seen by arc formations for small momenta (see also appendix \ref{app:pair} for a discussion of this structure). If the effective hopping changes its size, we see that the region of large transmission becomes overall broader for a larger effective hopping and the substructure becomes more evident. In contrast, for a smaller effective pair hopping a more narrow structure is found.
The central maxima of the pair transmission occur if $\cos(k)-\mu_p\approx 0$ due to the first term in the pair transmission Eq.\ref{equ:transmission_pair}, i.e if the energy of the incoming pair corresponds to the depth/height of the potential well. Depending on the momentum they occur at the negative or positive values of $\mu_p$, i.e. a well or a barrier. The side structures correspond to the conditions  $\cos(k)-\mu_p\approx \pm \gamma_p$ and stem from the second term in the pair transmission Eq.~\ref{equ:transmission_pair} (see also appendix \ref{app:pair} for a discussion of this structure).  This means that the energy of the incoming particle reduced by the energy of the effective potential matches the effective hopping amplitude. 

We will recognize this behaviour in the next section where we turn to the discussion of the dependence of the transmission on these physical parameters. 

\section{Discussion of the pair transmission}
\label{sec:pair_discussion}
\begin{figure*}
\setlength{\unitlength}{0.1\textwidth}
\includegraphics[width=.99\textwidth]{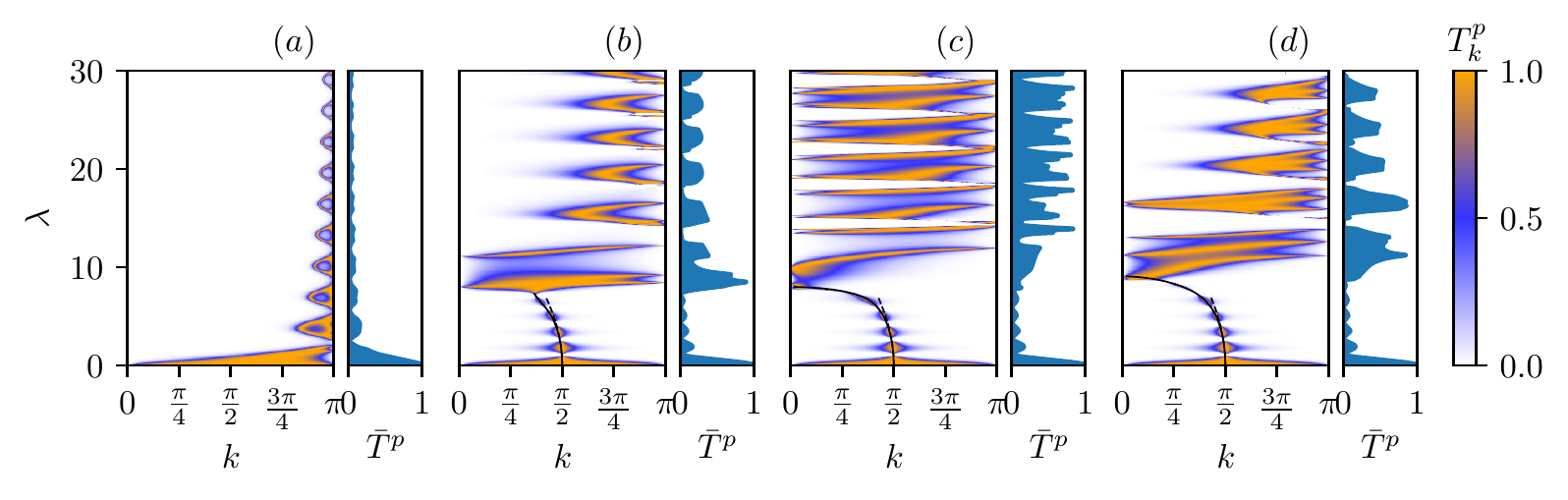}
\caption{Pair transmission as function of momentum $k$ and driving strength $\lambda$ for different interaction strength $|U|/\omega $ equal to (a) $0.5$, (b) $9.9$,(c) $10.1$, (d) $10.5$. The plot on the right side shows the momentum averaged transmission. For small $\lambda$ we also give the central transmitted momentum $k\upd{c} = \arccos(\mu\ind{p})$ (solid line) and its Taylor approximation (\ref{equ:kc}).}
\label{fig:pair_transmission_diffU}
\end{figure*}

In this section we discuss our results for the pair transmission and its dependence on the physical parameters of the system. 
In order to give an insight into the behaviour, we give a plot for different fixed interaction strength $|U|/\omega =0.5, 9.9, 10.1, 10.5$ in Fig. \ref{fig:pair_transmission_diffU}. 

The pair transmission has a complex structure. In general, as function of $\lambda$, three different regimes occur:\\

(i) For $\lambda \lesssim |U|/\omega$ the transmission has a sequence of maxima. For $\lambda=0$ the position of these maxima corresponds to $k=\pi/2$. 
In contrast to the maxima of the single particle transmission, the center of the momentum region at which the  maxima occur is not anymore the constant momentum $\pi/2$, but it shifts to lower momenta with increasing $\lambda$ forming a pronounced arc. The momentum region over which the maxima occur narrows with increasing $\lambda$ (see  Figs. \ref{fig:pair_transmission_diffU} b,c,d). In the regime of small $\lambda$ we can use the approximations (\ref{equ:gamma_taylor}) and (\ref{equ:mu_taylor}) for the functions $\gamma_p$ and $\mu_p$. Using these approximation one finds that the oscillation in the $\lambda$ direction mainly result from the oscillations of $\gamma_p$ or in this limit of the Bessel function $J_0(2\lambda)$. The transmission (see Eq.~\ref{equ:transmission_pair}) is proportional to $\gamma_p^4$ which leads to strong oscillations. In contrast, the position of the maxima which forms the arc sharp is due to the dependence of $\mu_p$ on $\lambda$. As discussed in Section \ref{sec:transmission_effective_pair_hamiltonian}, the pair transmission function Eq.~\ref{equ:transmission_pair_params} shows maxima if the condition $ \cos(k)-\mu_p\approx 0$ is fulfilled, i.e.~ if the energy of the incoming pair corresponds to the depth of the potential well. Note, that for small $\lambda$ the other maxima at  $ \cos(k)-\mu_p\approx \pm \gamma_p$ do not appear. Using the approximation $\mu_p\propto \lambda^2$, this leads to the arc seen at low values of $\lambda$, where the maximum transmission occurs at momentum $k=\arccos(\mu_p)\approx \arccos ( \frac{1}{|U|^2/\omega^2 - 1}\frac{\lambda^2}{2})$. This means that at low driving amplitude $\lambda \lesssim U/\omega$ the value of the momentum at which the pair transmission maximum occurs is shifted by increasing the effective potential $\mu_p$ of the pair model. The maximal value occurs if the energy of the incoming pair is equal to the depth of the well. In contrast, the amplitude of the transmission maxima mainly depend on the effective pair hopping amplitudes $\gamma_p$. 

(ii) For $\lambda > |U|/\omega$ we find alternating regions of high and low pair transmission.   For large $\lambda$, the possible values of the function $\gamma_p$ become smaller than 1. In contrast $\mu_p$ oscillates around the value $-1$ which corresponds to a large potential barrier. As for the low driving case, the overall structure of the oscillations of the transmission amplitude correspond mainly to the oscillations of the effective hopping amplitude $\gamma_p$ with $\lambda$, since $\gamma_p^4$ gives the prefactor of the pair transmission (Eq.~\ref{equ:transmission_pair}). Additionally to the overall oscillations with $\lambda$, the transmission shows a pronounced substructure which we are going to analyse in the following.

At large driving values $\lambda$, the maximal transmission is overall concentrated at higher momenta. This has its origin in the relation $ \cos(k)-\mu_p\approx 0, \pm \gamma_p$ for the position of the maxima of the pair transmission. Note that in contrast to the low $\lambda$ region, here the sign of the effective potential $\mu_p$ changed. Thus, the condition can be most easily fulfilled at momenta  between $k=\pi/2$ to $k=\pi$ where the cosine function becomes negative. A clear feature is also that the transmission is often organized in branches stemming from the energy matching conditions, in particular for very high or low momenta.

(iii) Another very pronounced feature is a high transmission around $\lambda \approx |U|/\omega$ throughout the entire Brillouin-zone. This very strong transmission is mainly due to the prefactor  $\gamma_p^4$, which exhibits a strong maximum around $U$. Within this transmission maxima around $\lambda \approx |U|/\omega$ a pronounced substructure occurs. Three strong branches of the pair transmission can be seen within this maximum of the transmission which are reaching over almost the entire momentum range.  

\section{Momentum filter}
\label{sec:filter}
The transmission properties of the single particles and pairs can be used to engineer different types of filters. In Ref.~\cite{HuebnerSheikhan2022} we discussed in particular the possibility to either block pairs of single particles and thus, separate the two components from each other. In the following we would like to focus on the possibility to not only filter out pairs or single particles, but also to prepare these with specific momenta. For only the single particles such filters have been discussed in \cite{ThubergReyes2017,ReyesEggert2017}.

\subsection{Filter specific momenta}

\begin{figure}[hbtp]
\centering
\includegraphics[width=.40\textwidth]{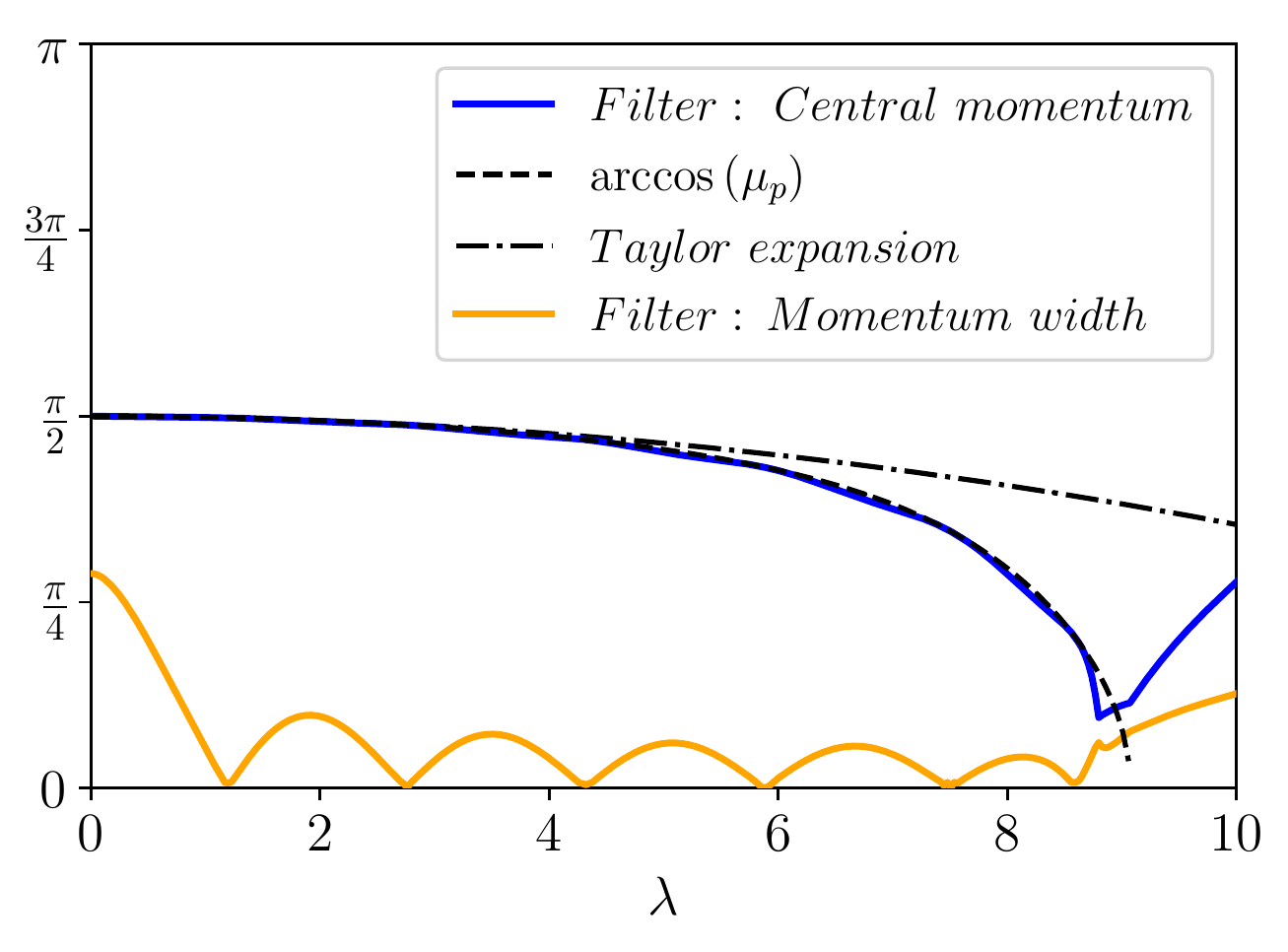}
\caption{Central momentum and momentum width as function of $\lambda$ for $|U|/\omega = 10.5$. For the central momentum we also give the expected curve given by the arcus cosine of $\mu\ind{p}$ and its Taylor expansion (dashed).}
\label{fig:momentum_pair_filter}
\end{figure}

In the discussion of the pair transmission (cf.~Fig.~\ref{fig:pair_transmission_diffU}) we have seen that for sufficiently large $U/\omega$ in the region of $\lambda < |U|/\omega$ the pair transmission shows a narrow peak at a specific momentum. The central momenta of this peak shifts along an arc with increasing $\lambda$ from $k=\pi/2$ towards smaller momenta. This maximum stems from the condition $\cos(k)-\mu_p\approx 0$ and results in a central momentum $k_c$ which is determined by $\mu\ind{p}$:

\begin{align}
k\upd{c} = \arccos(\mu\ind{p}) = \frac{\pi}{2}-\frac{1}{\frac{|U|^2}{\omega^2}-1}\frac{\lambda^2}{2} + \order{\lambda^4}
\label{equ:kc}
\end{align}
where we used the Taylor expansion (\ref{equ:mu_approx}) for small $\lambda$. In Fig~\ref{fig:momentum_pair_filter} we show a plot of the (numerically integrated) exact central momentum, together with (\ref{equ:kc}), and the momentum filter width, e.g.~the standard deviation obtained by normalizing $T\upd{p}_k$ to a probability distribution. 
We see that the filtered average momentum can be chosen by the parameters in the interval between a small value and $k=\pi/2$. The variance in the filtered momentum distribution is very small. This could be used in order to prepare a wave packet of a pair with a very well defined momentum. Let us note that one can at the same time choose the single particle transition to almost vanish, if one adjusts the interaction parameter instead of the driving strength $\lambda$.

\begin{figure}[hbtp]
\centering
\includegraphics[width=.45\textwidth]{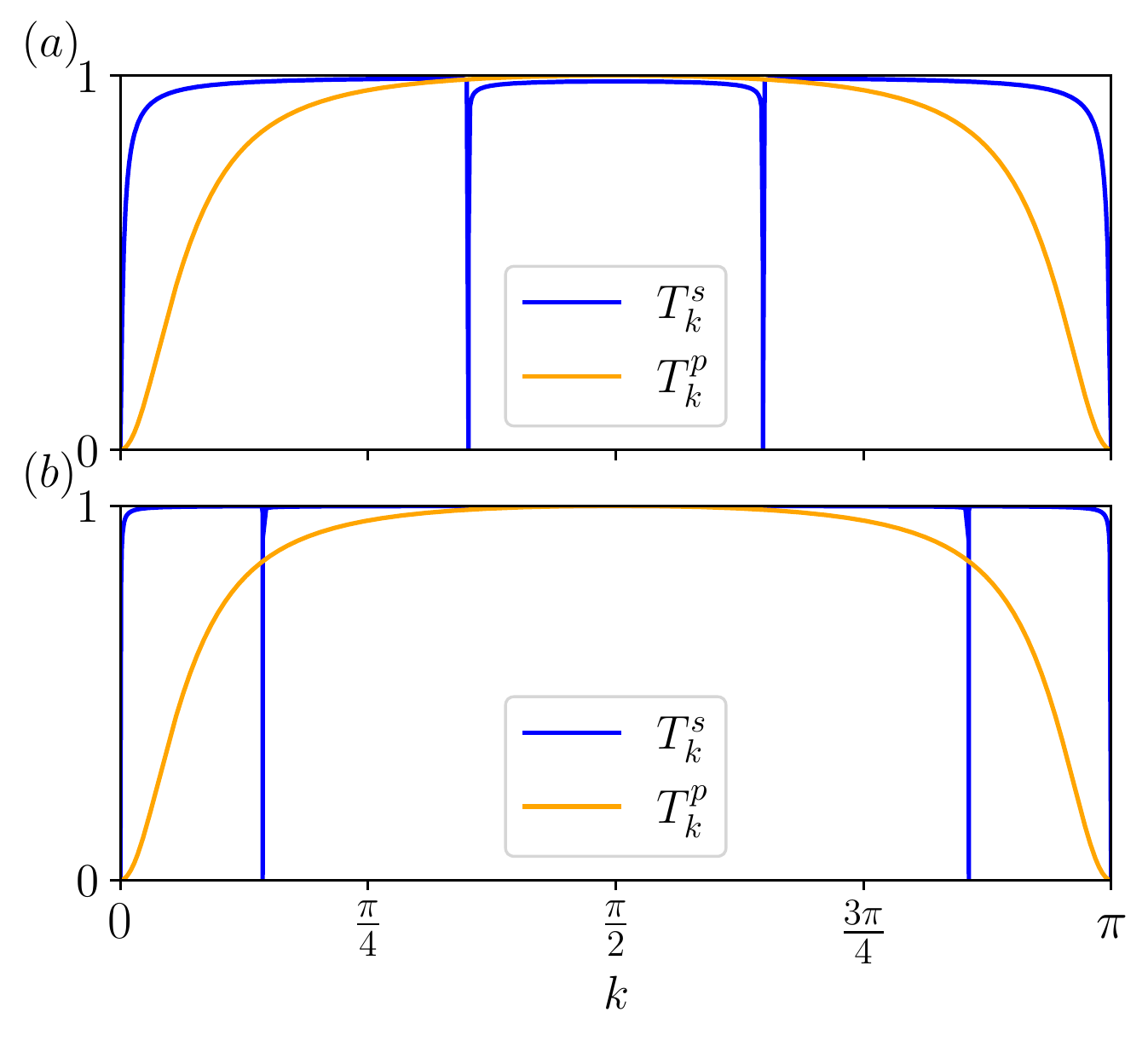}\\
\caption{Single particle transmission and pair transmission for $\vert U\vert/J = 100$, $\lambda = 0.3$ and (a) $\omega = 1.1 J$, (b) $\omega = 0.2 J$. Note that the pair transmission is the same in (a) and (b), since for the pairs the parameters lie in the large frequency limit.}
\label{fig:single_pair_largeU}
\end{figure}

One can also filter out single momentum modes of the pair and/or of the single particles, respectively. In order to do this, one can use the Fano resonance (section \ref{sec:pair_low}) which appears only at frequencies of the order of the respective hopping amplitude.  Let us note, that as the driving frequency has to be in order of the hopping amplitude, this effect needs the study of the full time-dependent scattering and cannot be found with effective models in the high frequency regime. The occurring Fano resonance is very narrow and allows to filter out momentum states in a very narrow region of the band both in the single particle and pair sector. In Fig.~\ref{fig:single_pair_largeU} it is shown how the Fano resonance can be used to filter out single particle states at a certain quasi momentum, while letting the transmission of the pairs undiminished. One can see that the location of this peak can be tuned by the driving frequency and follows Eq. \ref{eq:fano_single} for the single particle. Additionally one can also filter out a narrow momentum region in both the pair and the single particle mode. This had already been shown in Fig.~\ref{fig:singlepair_transmission_fano}, where one filters out certain momentum modes in both the single particle and the pair sector.  Also here the filtered momenta can be chosen by the parameters used. With the help of the interaction strength, the position of the resonance of the pair can be shifted independently from the position of the single particle Fano resonance. 

\section{Conclusion}
We have investigated the transmission through a single driven impurity in the Hubbard model. We have focused on the momentum resolved transmission through the periodically driven impurity of a pair of a spin up and a spin down fermion formed by the attractive interaction strength. We have discussed in detail the form of the momentum dependent transmission which goes beyond the results presented in our previous work \cite{HuebnerSheikhan2022}, where the total transmission was analyzed. We find a very complex behaviour of the pair transmission. In particular a broad transmission peak is present if $U/\omega \approx \lambda$. For $\lambda < U/\omega$, in contrast, only a  very narrow momentum window shows a sizable transmission. These complex structures can be used in order to engineer different filters targeting for example different momentum windows of the pairs. Whereas we could identify the transition in a wide region of the parameter space, the low frequency region could only be investigated in  the large interaction regime. A further investigation of the problem at small and intermediate interaction strength would be very interesting, but also challenging. 

\textit{Acknowledgments}
 We thank M. K\"ohl, H. Ott, I. Schneider, and A.-M. Visuri for stimulating discussions. We acknowledge funding from the Deutsche Forschungsgemeinschaft (DFG, German Research Foundation) in particular under project number 277625399 - TRR 185 (A5,B3) and project number 277146847 - CRC 1238 (C05) and under Germany’s Excellence Strategy – Cluster of Excellence Matter and Light for Quantum Computing (ML4Q) EXC 2004/1 – 390534769 and the European Research Council (ERC) under the Horizon 2020 research and innovation programme, grant agreement No.~648166 (Phonton).
 
\appendix
\section{Floquet-Schrieffer-Wolff transformation}
\label{app:fswt}
The Floquet-Schrieffer-Wolff transformation extends the ideas of the Schrieffer-Wolff transformation to Floquet-systems. The original derivation~\cite{BukovPolkovnikov2016} combines the idea of the Schrieffer-Wolff transformation (derivation of an effective Hamiltonian) with ideas from the high-frequency expansion (fast oscillating terms will average to zero). Here we would like to sketch a different approach which is based on mapping the time-dependent system onto an equivalent static system (extended Hilbert space) and then to use the standard Schrieffer-Wolff transformation~\cite{BravyiLoss2011} on that system to derive an effective Hamiltonian. 

Consider a Floquet-system given by a Hamiltonian $U\vb{F}+\alpha\omega\vb{G}(\omega t) = U\vb{F}+\alpha\omega\sum_l\vb{G}_le^{-il\omega t}$, where $\alpha \ll 1$. In our case $\alpha = \tfrac{J}{\omega}$ and
\begin{align}
	\vb{F} &= -1+\sum_n \vb{n}_{n\uparrow}\vb{n}_{n\downarrow}\\
	\vb{G}(\omega t) &= -\sum_{n\sigma} \qty(g_n(\omega t)\vb{c}_{n\sigma}^\dagger\vb{c}_{n+1\sigma} + \mathrm{h.c.})
\end{align}

Note that we added the offset $-1$ to $\vb{F}$ so that $\vb{F}$ vanishes on the all states which contain exactly one pair and no single particles. First let us change the time variable in the system to $\phi = \omega t$ which yields the Hamiltonian 
\begin{align}
	\vb{H} = \tfrac{U}{\omega}\vb{F}+\alpha\vb{G}(\phi) = \tfrac{U}{\omega}\vb{F}+\alpha\sum_l\vb{G}_le^{-il\phi}
\end{align}

The extended Hilbert space formalism is a well established way to map a time-dependent system onto a static system in one higher dimension (see for instance~\cite{EckardtAnisimovas2015}). The extended system can be described as follows: It consists of an infinite number of copies of the system labelled by an integer $l$ (one copy for each Fourier component in time). The copies $l$ and $l'$ are coupled by $\alpha\vb{G}_{l'-l}$. Additionally each copy $l$ has an energy offset $-l\omega$. Without perturbation ($\vb{G}_l = 0$) only the physical copy $l=0$ would be populated, but due to the coupling eigenstates also expand to the other layers. We interpret a change from copy $l$ to $l'$ as an absorption/emission of $l'-l$ energy quanta from/into the driving. Thus we can think of the copies as an convenient way to keep track of the current energy of the system.

We can now apply the Schrieffer-Wolff transformation to it. Define the projection $\vb{P}$ onto the subspace which describes exactly one pair and no single particles in copy $l=0$. By construction $\vb{F}\vb{P} = 0$. Also it is not hard to see that $\vb{P}\vb{G}_l\vb{P} = 0$ The Schrieffer-Wolff transformation applied to the extended Hilbert space produces an effective Hamiltonian on the subspace $\vb{P}$:
\begin{align}
	\vb{H}\upd{eff} = -\alpha^2 \sum_l \vb{P}\vb{G}_{-l}\frac{1}{\tfrac{U}{\omega}\vb{F}-l}\vb{G}_l\vb{P}
\end{align}

We can understand the formula as follows: If we have a static system $\vb{G}_l = 0$ for $l\neq 0$ the formula reduces to the usual Schrieffer-Wolff transformation $\vb{H}\upd{eff} = -\alpha^2 \vb{P}\vb{G}_{0}\frac{1}{\tfrac{U}{\omega}\vb{F}}\vb{G}_0\vb{P}$. The effective Hamiltonian acts as on a state in $\vb{P}$ by first applying $\vb{G}_0$ which takes the state outside of $\vb{P}$ and then applies $\vb{G}_0$ again to go back to a state in $\vb{P}$. In our case $\vb{G}_l$ is a hopping Hamiltonian, the first $\vb{G}_0$ moves one of the two particles (thereby breaking the pair for short time) and the second $\vb{G}_0$ then either moves that particle back or moves the other particle to the same site, such that we have again a pair. In the case of Floquet-driving we can do exactly the same, but now we also have the option that the first application $\vb{G}_l$ changes the copy to $l$ and the second $\vb{G}_{-l}$ changes it back to $l=0$. That is the reason for the appearance of the sum over $l$. However since all the $\vb{G}_l$ are still hopping Hamiltonians the result (\ref{equ:hamiltonian_pair}) is still a sum of a pair hopping, a pair interaction term and a potential term as in the usual Schrieffer-Wolff transformation. The only objects one has to calculate are the amplitudes in front of each term which depend on the details of the driving.

\section{Discussion of the structure of the transmission in the effective pair Hamiltonian}
\label{app:pair}
In order to analyze the features of the pair transmission (\ref{equ:transmission_pair}) we rewrite it as follows:

\begin{align}
T\upd{p}_k &=\frac{1}{1+\xi_k^2\varrho_k^2} \frac{1}{\xi_k^4 +\left(\tfrac{1}{\varrho_k^2}-2\right)\xi_k^2+1}
\label{equ:transmission_pair_params}
\end{align}
Here we defined $\xi_k = \frac{|\mu\ind{p}-\cos{k}|}{|\gamma_p|}$  and $\varrho_k = \frac{|\gamma\ind{p}|}{|\sin{k}|}$.
This representation of the pair transmission has the advantage that it only depends on two parameters, $\xi_k$ and $\varrho_k$.
The parameter $\xi_k$ can be interpreted as being proportional to the difference between the maximal value of the potential barrier ($2\mu_p<0$) or potential well ($2\mu_p>0$) and the energy of the incoming particle $2\cos(k)$ rescaled by the effective hopping amplitude $\gamma_p$. The value of the parameter $\varrho_k$ goes from $\gamma_p$ at $k=\pi/2$ to $\infty$ for $k\to 0,\pm \pi$.

\begin{figure}[!h]
	\centering
	\includegraphics[width=.48\textwidth]{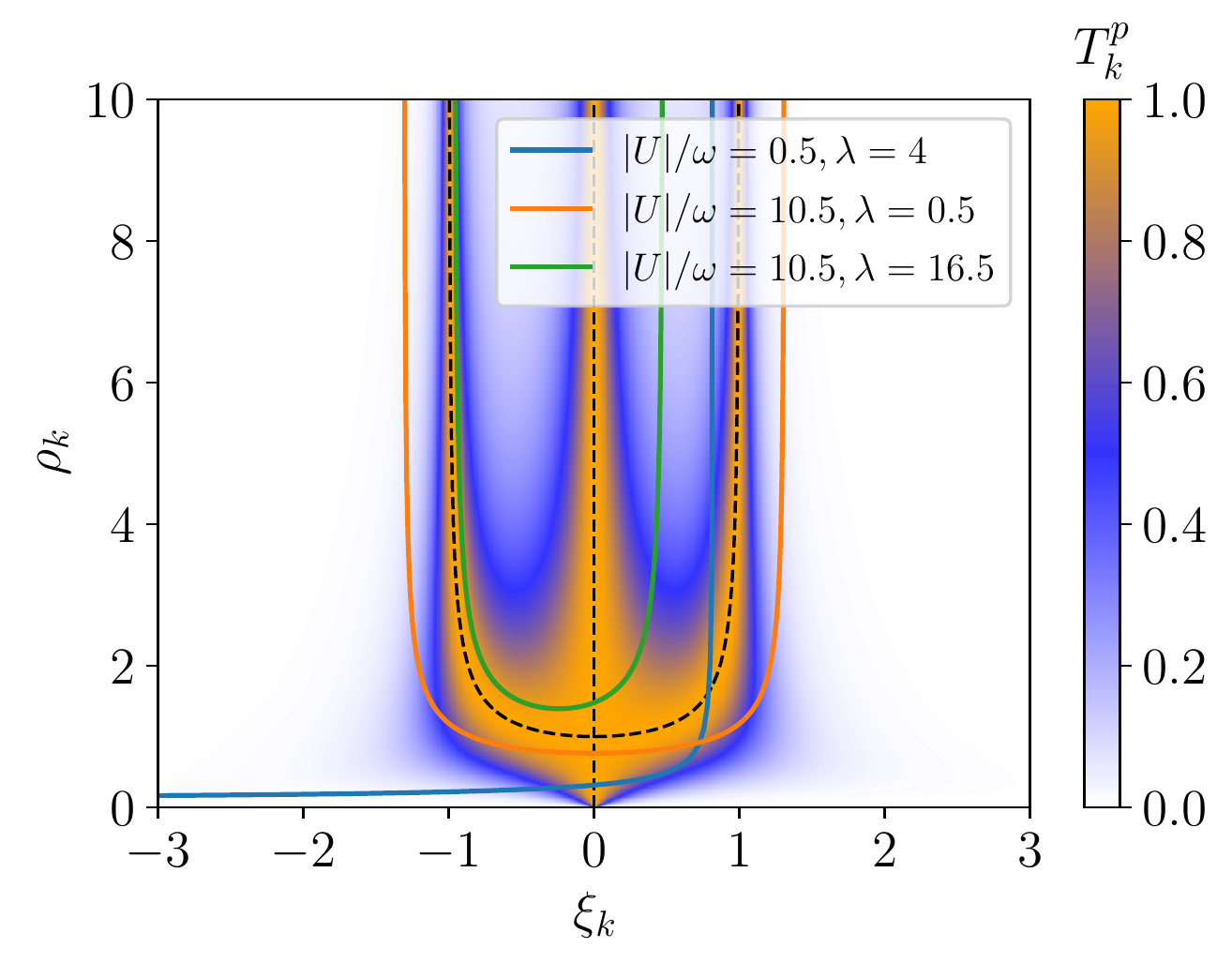}
	\caption{Pair transmission in the effective pair model (Eq.~\ref{equ:hamiltonian_pair}) as a function of $\xi_k$ and $\varrho_k$ as given by Eq.~\ref{equ:transmission_pair_params}. Dotted lines: Lines of maximal transmission $T^p_k = 1$.  The solid lines show how the two parameters $\xi_k$ and $\varrho_k$ vary by changing the momentum $k$ from $0$ to $\pi$ for different interaction strength $U$ and driving amplitude $\lambda$. }
	\label{fig:transmission_pair_params}
\end{figure}

A colorplot of the dependence of the transmission on these two parameters is shown in Fig.~\ref{fig:transmission_pair_params}. However, one should note that $\xi_k$ and $\varrho_k$ are depending both on $k$ and on the combination $U$, $\lambda$ and $\omega$ through the parameters $\gamma_p$ and $\mu_p$. Thus, for a fixed value of interaction $U$, the driving frequency $\omega$ and driving strength $\lambda$, only some values of $\xi_k$ and $\varrho_k$ are possible varying the momentum $k$.

The pair transmission function has a very particular form. At large values of $\varrho_k$ the transmission is usually small unless $|\xi_k| \approx 0,\pm 1$. This has its origin in the structure of the denominator ocurring in the expression for the pair transmission and is related to the triangular form of the potential step. If $\varrho_k^2>1/2$, the function $\xi_k^4 +\left(1/\varrho_k-2\right)\xi_k^2+1$ has two minima at $\xi_k\approx \pm \sqrt{1/2 (2-1/\varrho_k^2)}$. For large values  $\varrho_k$ the position of these minima become $\xi_k\approx \pm 1$. Together with the first contribution $\frac{1}{1+\xi_k^2\varrho_k^2}$ which has a maximum at $\xi_k=0$, this leads to the three pronounced maxima seen in Fig.~\ref{fig:transmission_pair_params} at large values of $\varrho_k$.

For smaller $\varrho_k$ the three maxima join to a single maximum around $\xi_k = 0$. The position of the maxima follows approximately the function $\xi_k\approx \pm \sqrt{1/2 (2-1/\varrho_k^2)}$ of the minima of the second denominator until for ${\varrho_k}^2\leq 1/2$ only one maximum exists at $\xi_k=0$. For small $\rho_k<1/2$ physically, this means that the transmission becomes maximal, in the situation where the energy of the incoming pair corresponds exactly to the minimal value of the potential triangle. 

Since the parameters $\xi_k$ and $\varrho_k$ are themselves complicated functions of the system's parameters, we discuss in the following section how the transmission function depends on the physical parameters.

\end{document}